# Neuromorphic computing for anomaly detection in a laser powder bed fusion process

Shreyan Banerjee¹, Aasifa Rounak¹, Cathal Hoare¹, Denis Dowling¹, Vikram Pakrashi,¹,

ᵃ*Dynamical Systems and Risk Laboratory, School of Mechanical and Materials Engineering, University College Dublin, Belfield, Dublin 4, Dublin, Ireland*
ᵇ*i-FORM - SFI Research Centre for Advanced Manufacturing, School of Mechanical and Materials Engineering, University College Dublin, Belfield, Dublin 4, Dublin, Ireland*

**Abstract**

This study is the first application of spiking neural networks(SNNs) for anomaly detection in the Laser Powder Bed Fusion (LPBF) additive manufacturing process. The neural networks were used to identify print processing anomalies generated by dropping of laser energy during the printing of individual layers in a Ti-6Al-4V alloy lattice structures. Associated changes in the laser generated melt pool were observed using an in-process photodiode monitoring technique. photodiode sensors capturing plasma and infrared radiations reflected from the print bed of the metal 3D printer were utilized to detect sudden changes caused by anomalies during the printing process. The algorithm is first implemented on non-neuromorphic hardware including a central processing unit (CPU), on Field Programmable Gate Arrays (FPGA) and then on neuromorphic Intel's Loihi chip. Improved detection of anomalies is achieved by adjusting the spike latency of the neural network, which reduces masking of information by noise within the monitored temporal signal. The work demonstrates the possibility of using low-power neuromorphic chips within an edge framework for anomaly detection in additive manufacturing and creates a framework for the process.

## 1. Introduction

Additive manufacturing (AM) has attracted substantial scientific and industrial attention over the past two decades, driven by advancements associated with Industry 4.0 and the increasing demand for customized, high-performance components. This technique of creating objects through layer-by-layer deposition of material based on a digital 3 dimensional (3D) model [1, 2, 3] often creates metal parts through laser powder bed fusion (LPBF) [4, 5] . Details of the process of LPBF can be obtained in Scime *et. al.* [6]. LPBF is widely used for load bearing components in aerospace, biomedical, and automotive industries and the detection of damage of flaws in the manufacturing process to ensure structural integrity is of significant interest. A list of critical flaws in this regard is also outlined in [7]. Post-build condition monitoring of components, often used in the industry, is neither cost nor time efficient, as defects in one layer get sealed in by the

*Email addresses:* shreyan.banerjee@ucdconnect.ie (Shreyan Banerjee), aasifa.rounak@ucd.ie (Aasifa Rounak), cathal.hoare@i-form.ie (Cathal Hoare), denis.dowling@ucd.ie (Denis Dowling), vikram.pakrashi@ucd.ie (Vikram Pakrashi,)

subsequent layers. Thus, real-time anomaly detection during a printing process is very important.

Various studies have been carried out for detecting anomalies in an additive manufacturing process using statistical analyses and machine learning tools. Doyle *et. al*'s work [5] uses the statistical "Search and TRace AnomalY" (STRAY) algorithm for detecting defects introduced by laser power reduction, in certain layers of a Ti-6Al-4V part, by sensing the plasma and infrared reflections off the print bed and utilising the Savitzky-Golay algorithm, the moving average, along with Gaussain and Butterworth smoothing techniques on the obtained time histories of the data over the deposited layers. Similarly, Mohammadi *et. al* [4] used acoustic emission (AE) data to detect cracks and porosities for LPBF 3D printing via K-means, principal component analysis (PCA), Gaussian mixture models, and variational auto-encoders. A review of supervised and unsupervised learning algorithms for anomaly detection in metal additive manufacturing (MAM) is available in Sahar *et. al* [8]. Geometry-informed nominal behaviour combined with PCA [9] usucessfully detected faults in LPBF process, as well as image processing techniques to identify various types of anomalies [6]. Thermal images of the melt pool of LPBF layers have also been analysed to identify regions of interest and to develop anomaly classifiers [10].

Despite these advances in statistical and machine learning for detecting anomalies of defects in LPBF processes, studies on real-time fault detection are few. To enhance both the processing efficiency and sustainability of AM processing, close to real-time feedback on the generation of printing anomalies such as porosity or wiper damage, is required. An edge computing framework, ensuring low power consumption while minimizing transmission and processing latency, can thus be useful. A promising solution in this regard is neuromorphic computing [11, 12], which processes data in a highly parallelized form using networks of interconnected spiking neurons processing data spatio-temporally. This approach also makes neuromorphic devices particularly suitable for anomaly detection in time series data. However, the low power and low latency benefits of spiking neural networks (SNNs) come with the cost of reduced precision. In anomaly detection applications, this limitation is not a critical concern, since the primary objective, very often, is to identify the occurrence of a sudden change rather than to measure its exact magnitude. Existing work in anomaly detection using SNNs indicate the possibility of this approach to be adapted for a range of sectors. Bauer *et. al*'s work [13] used recurrent spiking neural networks with reservoir computing to detect anomalies in ECG signals, using the DYNAP (Dynamic Neuromorphic Asynchronous Processor) chip, as did [14]. Chen *et. al* [15] used a multiscale event-based spatio-temporal descriptor to filter out anomalies from images recorded by a Dynamic Vision Sensor (DVS) and created a benchmark dataset for such anomalies. In a similar line, Annamalai *et. al*'s work [16] proposed a conditional Generative Adversarial Neural Network (cGAN) with deep learning to find anomalies in videos captured by an event-based camera using the time history of each pixel, creating another anomaly benchmark. Jaoudi *et. al*'s paper [17] demonstrated the use of an autoencoder-based SNN model and online learning for sensing anomalies to detect car hacking using Intel's Loihi chip. Joseph *et. al* [18] used spiking neurons to extract cepstral coefficients from time series signals for structural health monitoring via simulation using the Neural Engineering Framework (NEF) and on the Loihi chip. These successful applications indicate that SNNs can cater to real-time applications by adjusting the delays to match the latency of the sensor signals, process spatiotemporal data, and provide low power computation, which is an important criterion for standalone edge applications.

Despite these advantages, neuromorphic approaches are nascent in additive manufacturing, with the only notable work [19] using hyper-dimensional neuromorphic computing as a robust and energy-efficient tool to train a supervised learning model to detect anomalies in the printed layers. This method of anomaly detection using neuromorphic frameworks and hardware is of prime importance for an L-PBF process, as it provides the scope for online anomaly detection at a very low power budget. To address this gap, this works demonstrates one such example using the Nengo-Loihi backend.

This work investigates the use of the spiking neural network approach for anomaly detection. The study was carried out using in-process photodiode data obtained during the LPBF printing of Ti-6Al-4V alloy parts conducted by Doyle *et. al* [5]. In this study infrared and plasma photodiode data from the meltpool, under both normal LPBF processing conditions (termed 'healthy' throughout the text), as well as when a print anomaly (termed 'defective') occurred. Nengo, built [20, 21] was used to create an ensemble of spiking neurons to filter and convert time series data recorded by the sensors into spikes [5, 3] and subsequently detect the presence, location and magnitude of manufacturing defects corresponding to sudden changes in features of measured data. The approach was first tested on a non-neuromorphic hardware including central processing unit (CPU) and Field Programmable Gate Arrays (FPGA) and then on Intel's Loihi chip [22]. The obtained results were analyzed and compared.

In this paper, Section 2 discusses the underlying concepts of spiking neurons and NEF in detail, followed by a brief description of the LPBF process examined in this study. This is followed by a discussion on the methodologies used, both hardware and software. Section 3 discusses the results obtained after implementing the anomaly detection algorithm on CPU, FPGA, and Loihi neuromorphic board. Finally, Section 4 presents the conclusions of the study, discusses the shortcomings of the proposed methodology and highlights some future directions.

## 2. Materials and Methods

### 2.1. Tools and Concepts

This work utilizes Nengo [20] to process temporal data collected from the photodiode sensors deployed in the LPBF process to detect defects while printing. A brief review of the conceptual details of the tool used are discussed in the following sections.

#### 2.1.1. Spiking Neurons: Leaky-Integrate-and-Fire Model

Spiking neurons compute data spatio-temporally in the form of spikes of electrical voltage or current. This is a mixed signal form of computation, where the input to a synapse and output from a neuron are digital in the form of spikes and the processing within the neuron is analog. In this work, the Leaky-Integrate-and-Fire (LIF) model is used because it is computationally relatively inexpensive and popular in other applications [23, 24]. The dynamics of a LIF neuron, are presented as Equations 1 and 2.

$$\tau \frac{dV}{dt} = -(V - E_l) + \frac{I}{g_L},\qquad(1)$$

$$I_{out} = \begin{cases} I_{spk} & \cdots V \geq V_{th} \\ 0 & \cdots otherwise. \end{cases}\qquad(2)$$

Here, $V$ stands for membrane potential, $E_L$ is the leakage voltage, $g_L$ is the leakage conductance, and $I$ is the synaptic current. $I_{out}$ is the output current with value $I_{spk}$ when $V$ crosses $V_{th}$, the threshold voltage. Neurons receive electrical current in spikes, producing an analog synaptic current after being low pass filtered at the synapse. This current leads to an increase in the membrane potential. Once the membrane potential surpasses a predefined threshold, a spike of electrical current is emitted at the output. This model is termed LIF, because of the leakage term $E_l$ which causes the membrane potential to decay when there is no input current. Each neuron, when used individually, is sensitive to only one part of the signal, known as the neuron's receptive field. The 'radius' of the ensemble in the NEF framework governs the receptive field of the spiking neurons in an ensemble. A collection of neurons with different receptive fields, as shown in Figure 1(a) can collectively span the entire measured signal under consideration. The process of synaptic filtering is mathematically expressed in Equation 3.

$$I(t) = \sum_{x=-\infty}^{\infty} I_{in}(x) e^{-(\frac{t-x}{\tau})}, \tag{3}$$

where $I_{in}$ is the spike train of current entering the synapse, and $\tau$ is the synaptic time constant. In this work a set of neurons designed in Nengo NEF [20] is used.

*2.1.2. Nengo*

Nengo is a versatile simulation framework that leverages spiking neurons and their interconnections to represent signals and simulate dynamic processes [21]. Rooted in the principles of the Neural Engineering Framework (NEF) [25], Nengo provides a mathematical foundation for translating complex dynamical systems into neural architectures. This approach not only affords a more biologically realistic representation compared to traditional rate-coded models but also enables the exploration of temporal dynamics that are critical for understanding processes such as synaptic plasticity and neural adaptation. Further research has extended these ideas by demonstrating Nengo's scalability and adaptability. For instance, studies have highlighted its compatibility with neuromorphic hardware platforms, making it a valuable tool for real-time control applications and large-scale neural simulations [26]. By simulating the precise timing of spikes, Nengo allows utilization of process information in a time-dependent manner, shedding light on the computational advantages of spiking networks. Nengo distributes the input signal vector **x** to be represented in a group of neurons with their corresponding activities $a$. Considering that each neuron has its own encoding vector $\mathbf{e}_i$, gain term $\alpha_i$ and constant background bias current $b_i$, the relationship between neural activity and input can be given by Equation 4.

$$a_i = G(\alpha_i \mathbf{e}_i \cdot \mathbf{x} + b_i). \tag{4}$$

Here, $G$ denotes the neural non-linearity function. This nonlinearity can be mathematically expressed in various forms, such as a sigmoidal function or the spiking LIF model (which has been employed in this work), etc. The relationship between neural activity and the input is known as tuning curves (see Figure 1 (a)). Each collection of neurons, called an ensemble, has its own radius, which decides the range of values it can represent. The tuning curve for each neuron is adjusted accordingly so that the ensemble can represent the full range of the input signal.

*2.1.3. Anomalies in 3D Printing*

The Ti-6Al-4V printing study was carried out using a Renishaw 500M Laser Powder Bed Fusion system, which is shown schematically in Figure 2 [5, 3]. This system incorporates an InfiniAM photodiode system for in-situ melt pool monitoring [27]. The laser specifications and powder particle diameter and type used is described in detail in Doyle *et. al* [5]. A $500W$ laser ($\lambda = 1.07 \mu m$), with a focused spot size diameter of approximately $80 \mu m$ was used. The test pieces were printed using Ti-6Al-4V grade 23 powder with powder particle diameters in the $15$–$45 \mu m$ range. Before the build process began, the chamber was vacuumed to remove any oxygen in it. Argon gas was then introduced to create an inert atmosphere. The build platform was maintained at a temperature of 170 degrees Celcius throughout the process. During the build, a photodiode was used to provide feedback on the laser energy input, known as the Beam Dump (BD) signal. Two more photodiodes (PD1 for plasma and PD2 for infrared emissions) were used to monitor emissions from the melt pool during the melting process. The InfiniAM (a process monitoring system) software was used to record the signals and convert them into bits. Print defects were artificially introduced into the alloy samples by systematically reducing the laser power for up to 9 layers during the printing of lattice structures. A direct corelation was obtained between the level of laser power reduction and the photodiode data obtained from the corresponding layer melt pools. The averaged data obtained from the InfiniAM software across each layer was passed through a network of spiking neurons. Upon passing, the high amplitude spikes, corresponding to the emissions recorded by the photodiodes when the nozzle deposits metal to an already molten pool from the previous pass, were suppressed from the measured time series signal to avoid these spikes being picked up as anomalies.

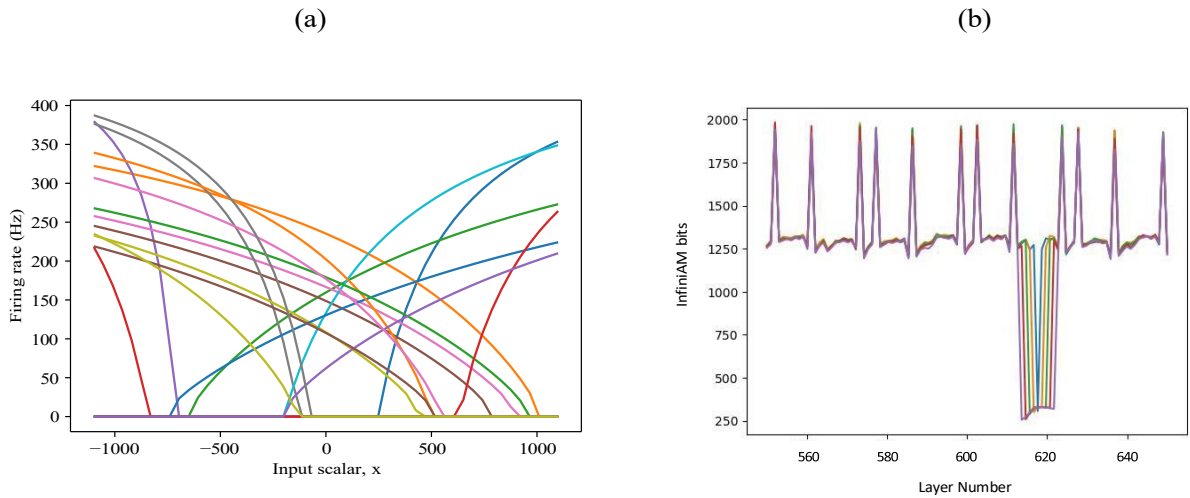

Figure 1: (a) Tuning curves showing the relationship between firing frequency and the input signal, (b) The raw mean plasma radiation signal recorded by the PD1 sensor for 66% reduction in source laser power.

*2.1.4. Data acquisition*

The InfiniAM system converts the voltage measurements from the photodiodes into digital bits containing data from the three installed sensors, captured at an acquisition rate of a $100 kHz$. For every layer of data, the mean voltage level in each photodiode was computed for further analysis, as described in the previous section. Three batches of lattice specimens were created, each incorporating several layers printed with an artificially reduced laser power. In the first batch, the energy of the laser was reduced by 33 %. The second and third batch of samples were manufactured with some layers deposited at a reduced energy level of 66 % and 100 % , respectively. For a 100% reduction in power, the laser did not fire during the deposition of the layers under consideration. The number of layers with reduced laser energy varied between 1, 3, 5, 7 and 9. The reduction in laser energy occurred between the deposited layers of 613 and 621. Figure 1 (b) shows the raw data plot for the reduction in laser power 66%, recorded by the PD1 sensor. The data is noticeably noisy and requires filtering. Furthermore, since the dataset analyzed corresponds to the additive manufacturing of lattice structures, at the junction of two adjacent rectangular lattices, a melt pool forms with a significantly larger volume of molten metal compared to that observed along the sidewalls of individual rectangles, see Figure 3. This increased quantity of molten metal resulted in an elevated intensity of optical emission spectral lines, leading to the periodic occurrence of these positive spikes as the lattice structure was fabricated. Thus, positive spikes in the data were observed to be unrelated to any significant phenomena associated with anomalous material deposition. Consequently, these spikes were clipped to enhance the algorithm's ability to accurately detect dips in laser intensity. The signal acquired from the InfiniAM software was passed through an ensemble of spiking neurons with appropriately chosen radius and synaptic filters to enable the clipping of the positive peaks and removing the high-frequency components in the signal respectively.

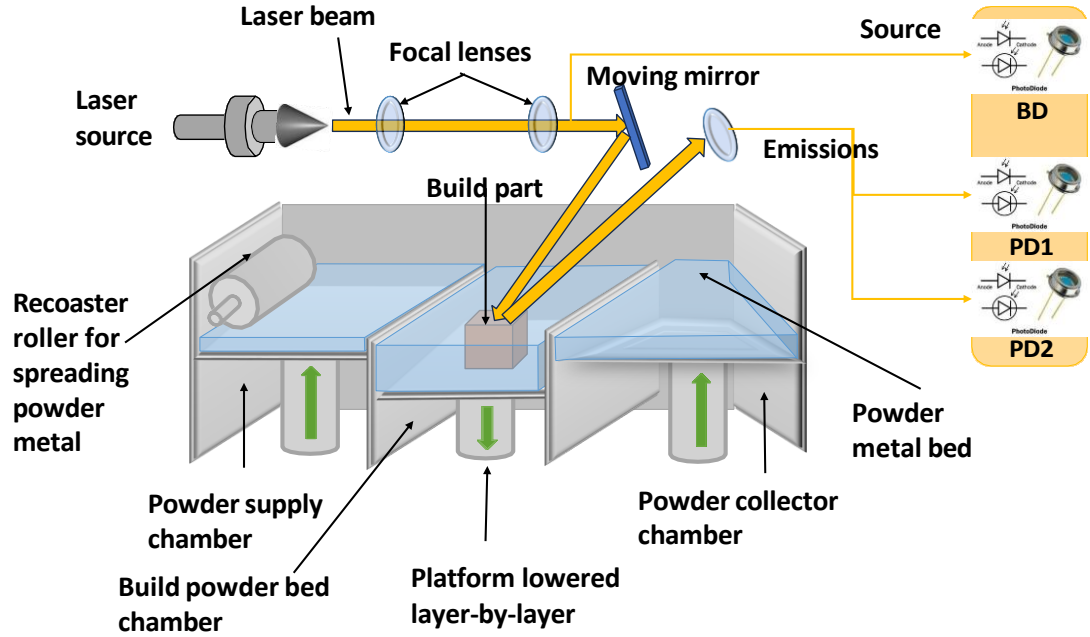

Figure 2: Schematic of the L-PBF process

*2.2. Methodology*

A series of simulations were conducted, followed by implementations on the CPU, an FPGA board and Intel's Loihi neuromorphic chip, as detailed below.

*2.2.1. Simulations*

The mean data on each layer was spike encoded using population encoding, which is default in NEF. These simulations are carried out using a Python integrated development environment (IDE), running on a 64-bit Ubuntu 22.04.4 LTS OS on a machine with an Intel® Core™ i7-8700K CPU @ 3.70GHz × 12, with a 32 GB RAM. A Nengo neural ensemble encoded the fed-in sensor data via an input node. The filtered and clipped data is decoded into a processed signal and sent to an output node. Table 1 shows a list of values for various parameters used in the network. The data corresponding to the defective print layers and that corresponding to the healthy printed layers are fed separately to the ensemble of neurons with the same parameters. The difference between the output signals in the two cases is then computed to find the percentage deviation. An alternate approach that potentially could have been implemented is to send both the defective and healthy data simultaneously through two input connections, the latter being inhibitory *i.e.* having a weight of −1. However, that would remove the DC component of the signal before it is fed to an ensemble. This can cause issues since the ensemble radius is tuned to clip off large positive spikes so that the signal dip due to the defect is prominent. However, if the DC component is removed, the signal trend becomes symmetrical about 0, and the ensemble would clip the signal from both the positive and negative regions. A practical approach that was adapted involves using two separate ensembles to independently filter and clip the data for both the damaged and undamaged states, and then subtract the filtered undamaged baseline from the filtered damaged data.

Figure 3: SEM of the printed Ti-6Al-4V lattice structures. The top images demonstrate that increasing the laser power and exposure time results in an increase in the diameter of the struts. The bottom images, from left to right, demonstrate the reduction in strut diameter with an increase in the laser beam spot size. Here $\omega$ the laser beam radius. The increased volume of molten metal at the junction between the struts leads to an enhanced level of optical emission intensity, resulting in periodic positive spikes in the spectral data.

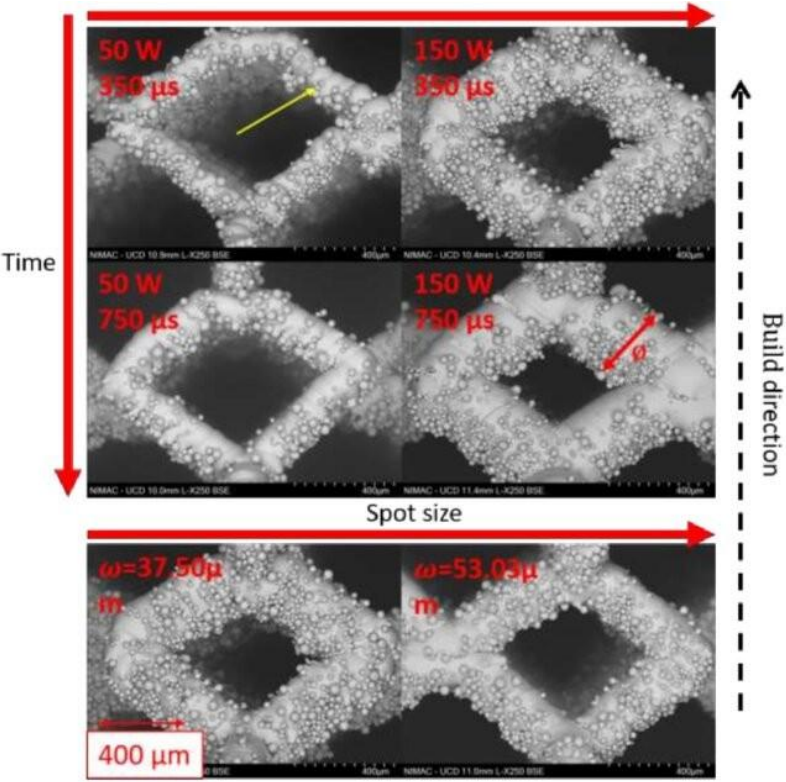

Table 1: Parameter Values used in the entire simulation

| Component | | |
|---|---|---|
| Name | Description | Value |
| Neurons | Number of neurons in the ensemble | 500 |
| Radius | Range of values that the ensemble can represent | 1100 |
| $dt$ | Simulation time step in Nengo | $0.001s$ |
| Presentation time | Time for which each data point is fed to the ensemble | $0.01s$ |
| Layer window | Layers for which the sensor data is fed into the network | 570 to 650 |
| Defective layers | Layers between which defects are introduced | 613 to 621 |

Table 2: Time constants for the network for processing different datasets on CPU

| | Component | |
|---|---|---|
| Sensor | Power reduction(%) | Time constant (s) |
| PD1 | 33 | 0.003 |
| | 66 | 0.002 |
| | 100 | 0.002 |
| PD2 | 33 | 0.006 |
| | 66 | 0.004 |
| | 100 | 0.004 |
| BD | 33 | 0.006 |
| | 66 | 0.006 |
| | 100 | 0.005 |

Figure 4 shows a schematic of the neural network used. To enable appropriate filtering of the input signal, the time constants of the input and output connections are chosen such that the low amplitude fluctuations due to noise are filtered out, retaining only the sudden changes in the signal due to the presence of anomalies. The time constant for the test cases are presented in Table 2. Time constants for the input connection and output connections from the ensemble *ens1* are chosen to be identical for this set of simulations.

*2.2.2. Hardware*

The simulations were repeated on a Field Programmable Gate Array (FPGA) board and Intel's Loihi neuromorphic chip. The Nengo-FPGA backend was used in this work to detect sudden changes in the 3D-printing process. This approach constitutes the implementation of spiking neural network on non-neuromorphic hardware. FPGA is a configurable hardware for testing prototypes and algorithms [28, 29, 30]. They are reprogrammable, allowing them to be used and rewired varied applications and can leverage the advantages of neuromorphic architectures due to their low latency and parallel processing abilities [30]. For this work, a Xilinx PYNQ-Z1 FPGA [28] from Digilent is used. Table 3 lists all the time constants used for each dataset.

The proposed detection scheme is subsequently run on the Loihi neuromorphic processor [22, 31]. Loihi, a digital neuromorphic board developed by Intel Neuromorphic Research Community (INRC), features a manycore mesh comprising of 128 neuromorphic cores, three embedded x86 processor cores, and off-chip communication interfaces that hierarchically extend the mesh in four planar directions to other chips. Each neuromorphic core implements 1,024 primitive spiking neural units (compartments) grouped into sets of trees constituting neurons [32]. Loihi was accessed remotely through Intel vlabs virtual machine that connects to a remote host running the code on a Loihi chip. Table 4 tabulates the time constants used in Loihi for processing different datasets. The number of neurons or the ensemble radius is kept the same for Nengo implemented on CPU, FPGA, and Loihi. Number of neurons chosen is 500 and radius is 1100.

Figure 4: The Nengo network model used to process the sensor data. $\tau$ represents the synaptic time constant for the input and output connections from the ensembles.

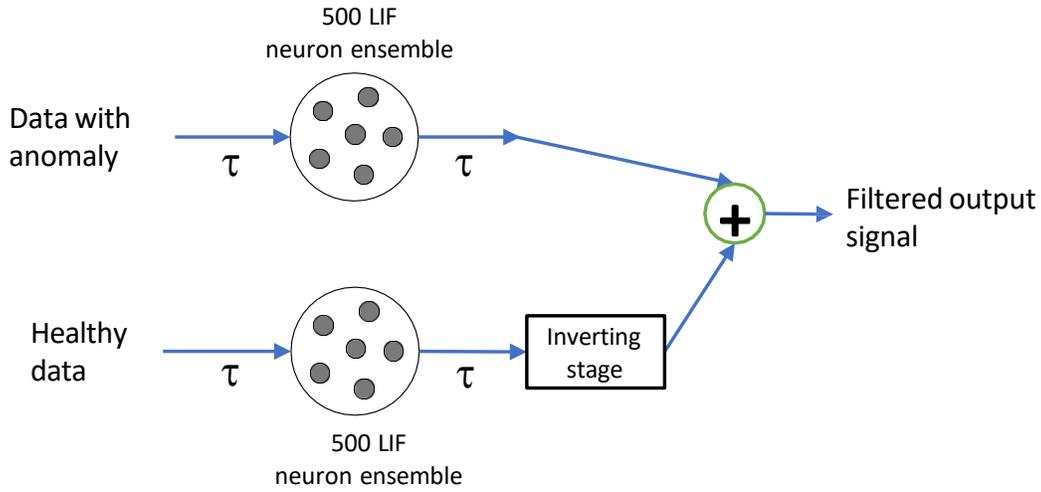

Table 3: Time constants for the network for processing different datasets on FPGA

| Sensor | Component Power reduction(%) | Time constant (s) |
|---|---|---|
| PD1 | 33 | 0.003 |
|  | 66 | 0.002 |
|  | 100 | 0.002 |
| PD2 | 33 | 0.008 |
|  | 66 | 0.004 |
|  | 100 | 0.004 |
| BD | 33 | 0.004 |
|  | 66 | 0.004 |
|  | 100 | 0.004 |

## 3. Results and Discussion

For both simulation and hardware, the sensor data filtered by the SNN ensemble is plotted for four different sensor-patch locations, each comprising of all 3 sensors mentioned above. The sensors are placed in bottom-left, bottom-right, top-left, and top-right (see Figure 5, 8, and 9), respectively. Here, results for only the bottom-left sensor patch is illustrated, as the others have similar characteristics which are described below. Each plot illustrates the percentage deviation between defective and healthy process data after being filtered by the SNN. Defects are introduced in 1, 3, 5, 7, and 9 layers between layers number 613 and 621. The results are plotted for a window of layers spanning between 570 to 650 so that the defective layers are presented. The results considered for 33%, 66%, and 100% reduction in the source laser power, for sensors BD, PD1, and PD2. It is observed that the most sensitive anomaly detection was possible using plasma photodiode measurements from the bed, as recorded by the PD1 sensor. Infrared recordings from the PD2 sensor are quite noisy even post filtering.

Table 4: Time constants for the Loihi network for processing different datasets

| Sensor | Component Power reduction(%) | Time constant (s) |
|---|---|---|
| PD1 | 33 | 0.003 |
|  | 66 | 0.002 |
|  | 100 | 0.002 |
| PD2 | 33 | 0.009 |
|  | 66 | 0.008 |
|  | 100 | 0.006 |
| BD | 33 | 0.02 |
|  | 66 | 0.008 |
|  | 100 | 0.006 |

Figure 5: Plot of the percentage deviation between the sensor data corresponding to the defective process as compared to the healthy process, for a sensor patch installed at the bottom-left end, after being filtered by the Nengo SNN ensemble running on CPU.

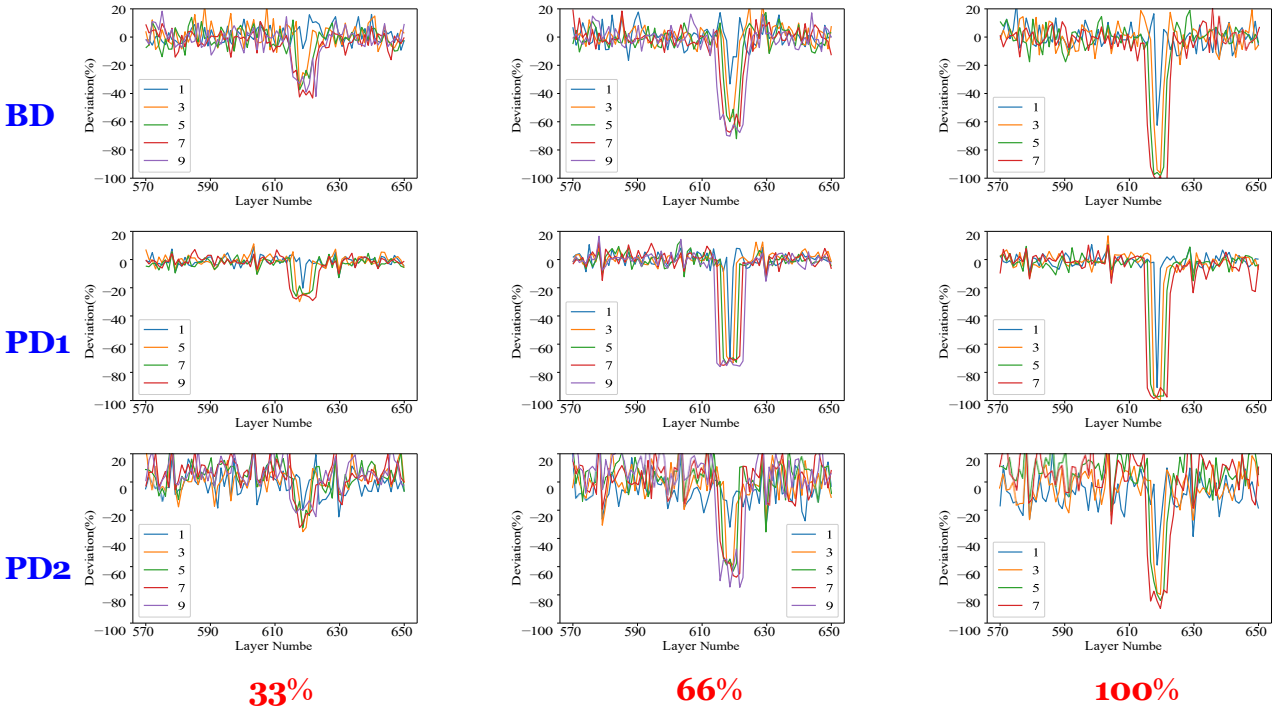

A number of observations are made from the simulations. Increasing the number of neurons in the ensemble increases precision. An ensemble with 500 neurons is observed to reasonably capture the signal characteristics, providing appropriate trade off between the resolution of the signal representation by the ensemble and its size. Increasing the synaptic time constant (from 0.001 s

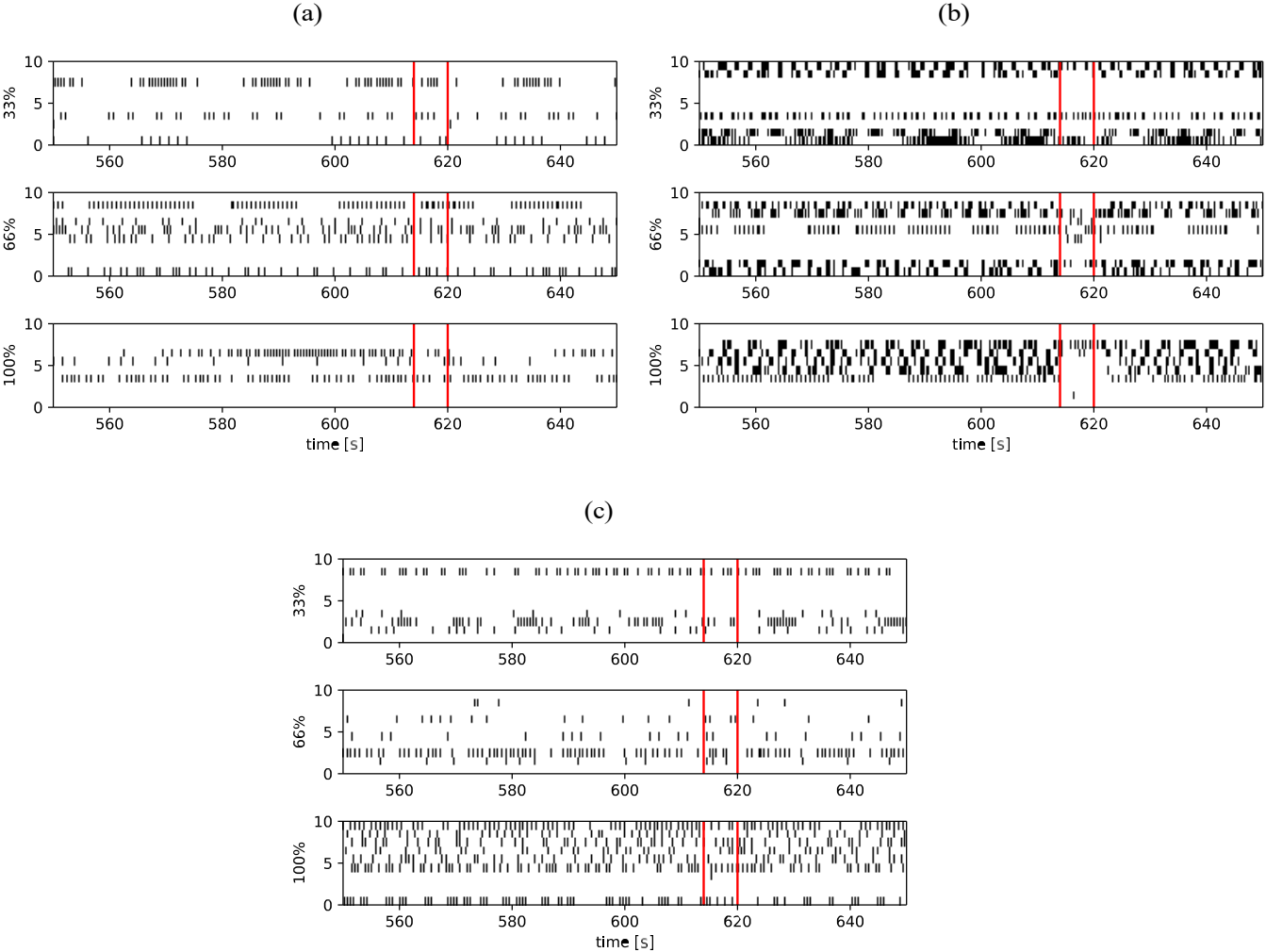

Figure 6: Spike raster plot for the neurons in the SNN ensemble when sensor data corresponding to the defective process is input for the sensor patch installed at the bottom-left end. The defect introduced by laser power reduction is over 7 layers and observed by (a) BD, (b) PD1 and (c) PD2 sensors. The introduction of defects is indicated by the layers between the red vertical lines.

Figure 7: F1 scores across varying time constants for anomaly detection using Nengo.

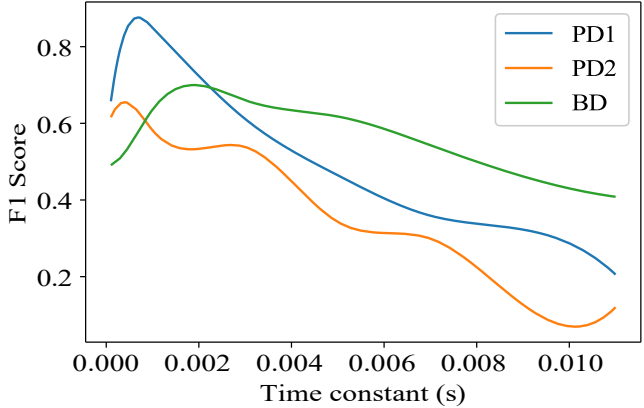

Table 5: F1 scores for the sensors at different locations

| Location | PD1 | PD2 | BD |
|---|---|---|---|
| BL | 0.710 | 0.500 | 0.666 |
| BR | 0.868 | 0.806 | 0.569 |
| TL | 0.836 | 0.614 | 0.688 |
| TR | 0.748 | 0.736 | 0.580 |

Table 6: F1 scores obtained by using SNNs compared to other filtering techniques.

| Filter | PD1 | PD2 | BD |
|---|---|---|---|
| Savitzky-Golay | 0.68 | 0.74 | 0.38 |
| Butterworth | 0.65 | 0.58 | 0.42 |
| Moving Average | 0.84 | 0.84 | 0.45 |
| Gaussian | 1 | – | 0.82 |
| SNN | 0.79 | 0.665 | 0.63 |

to 0.1 s) smoothens the signal, but a high time constant ($\approx 10^{-2}$s) suppresses the sudden change from the filtered signal for defects introduced in 1 and 3 layers. A high time constant also delays the occurrence of the sudden change in unfiltered signal, resulting in inaccurate estimation of the exact layer where the defect occurred (the estimated layers are pushed to a higher number than the layers where the defect actually occurred). Thus, the time constant must be chosen carefully. Compared to PD1, a higher value of the synaptic time constant for PD2 and BD data are chosen (Table 2) as the noise in these signals is higher. Here is a more polished and scientific rewording:

A lower time constant may cause defects resulting from a 33% reduction in laser power to go undetected, as the reduced signal-to-noise ratio impairs the ability to distinguish anomalies across successive layers. A higher time constant is necessary in this case as compared to the 66%, and 100% reduction cases, leading to a higher level of noise filtering. Choosing a suitable radius of the ensemble (here 1100), helps restrict the receptive field and clips off high amplitude positive spikes from the signal, focusing only on the troughs created due to the anomaly. The spike rasters generated by the neurons in the filtering ensemble, when fed with the sensor data corresponding to the defect (Figure 6) shows that the spikes get rarefied at the regions of damage, which is most prominently seen in the PD1 rasters. This is because few neurons spike only at the region of the defect, which tells us that these neurons encode the part of the signal representing the anomaly. F1 scores across varying time constants (Figure 7) indicate that for PD1, the maximum F1 score is obtained for a low time constant around 0.001. This is because PD1 signal is the cleanest, and filtering it with a higher time constant leads to a greater number of false negative observations. The F1 scores for PD2 show multiple peaks at different locations. This is because the signal is very noisy and at different values of time constants, where the filtering is adequate, the noise is removed to some extent and the false positives are eliminated. The F1 scores for BD show a maximum peak for a higher time constant around 0.004. This is again because the signal for BD is noisier and needs more filtering. The F1 scores for very low time constants of the order of $10^{-4}$ are lower because, with very low filtering, strong noise spikes are detected as defects, resulting in a lot of false positive observations. Table 5 shows the F1 scores for sensors placed at different locations in the setup. It is seen that PD1 sensors capture the LPBF process across layers reasonably well irrespective of the sensor location. But BD shows higher F1 scores for the left positions and PD2 records them for the right positions. This intuitively gives an idea about where to place the sensors, for capturing the process data with maximum accuracy. In this case, the BD sensor should be placed at either the bottom-left or the top-left patch, and the PD2 sensor should be spatially located at the top-right or the bottom-right patch. Table 6 compares the F1 scores obtained for the SNN-based filtering with those from other filtering techniques reported in [5]. It is seen that SNN provides a higher F1 score than Savitzky-Golay and Butterworth filtering techniques but provides a poorer F1 score than moving average or Gaussian techniques of filtering.

Figure 8: Percentage deviation between the sensor data corresponding to the defective process and the healthy process for the sensor patch installed at the bottom-left end after being filtered by the Nengo SNN ensemble running on an FPGA.

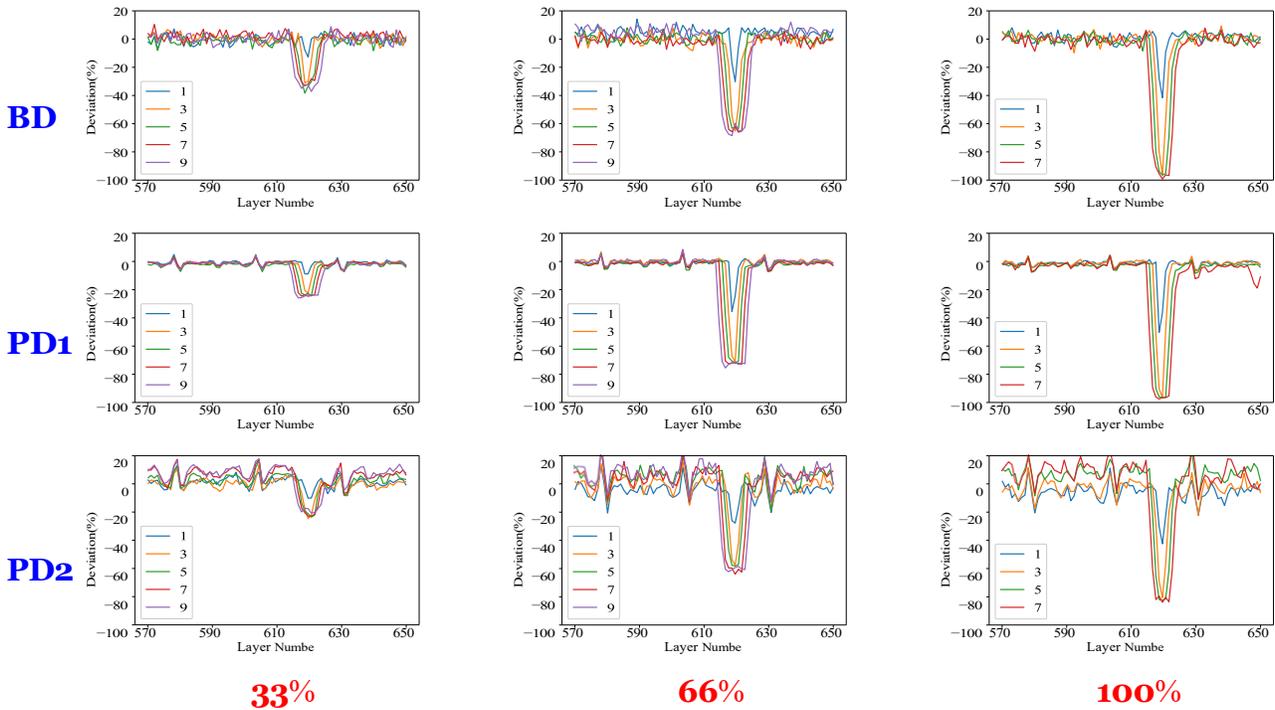

The Nengo FPGA board only allows ReLU and spiking-ReLU type neurons to be implemented. Using the default non-spiking ReLU neurons does not work, because the concepts of radius tuning and synaptic filtering apply only in the spiking domain. Consequently the signals are noisy with positive spikes. For a spiking ReLU ensemble, changing the radius does not show a pronounced effect on the range of inputs that is represented by the ensemble. A two-stage hybrid CPU-FPGA method can circumvent the problems described above. An ensemble of LIF neurons is first run on a CPU using radius-based clipping and synaptic filtering. In addition, another ensemble of spiking-ReLU neurons is run on an FPGA adding another stage of synaptic filtering. Each ensemble has 250 neurons so that the total number of neurons is consistent with the simulation. The synaptic time constant for processing the data with higher fluctuations corresponding to the 33% power reduction data is chosen to be larger as compared to the other power reduction cases, as seen in Table 3. It can also be observed that synaptic time constants for PD2 and BD data are chosen to be higher than that of PD1, to filter out the higher levels of noise in the PD2 and BD sensor recordings.

The Loihi implementation indicates that the fluctuations in the percentage deviation are much higher on the Loihi board as compared to the Nengo simulations on CPU. Increasing the synaptic time constant smoothens out the signal, but a very high time constant (of the order of $10^{-2}$s) eliminates the the sudden changes observed in layers 1 and 3. As observed for the CPU, a high time constant causes a delayed representation of sudden change in the filtered data, which results in an inaccurate estimation of the exact layer where the defect occurred, which is undesirable. Thus, a time constant in the range 0.002-0.02 has been chosen to trade off between defect detection and noise filtering. An alternative approach involves selecting the optimal time constant at which both false positives and false negatives are minimized. This has been observed to correspond to the maximum achievable F1 score, reflecting the best balance between precision and recall. Table 4 shows that the synaptic time constant for PD2 and BD data are higher than PD1 to account for the irregularities in the deviation between layers. For the same reason, the time constants for Loihi are chosen to be higher than the corresponding ones for CPU. From the plots, it is seen that the best representation of the anomaly can be found using the plasma reflections from the bed, recorded by the PD1 sensor. The infrared recordings from the PD2 sensor remain significantly noisy even after the application of filtering techniques. Higher noise adds more false positive observations, thus reducing the F1 score.

Figure 9: Percentage deviation between the sensor data corresponding to the defective process and the sensor data corresponding to the healthy process for the sensor patch installed at the bottom-left end after being filtered by the Nengo SNN ensemble running on Loihi.

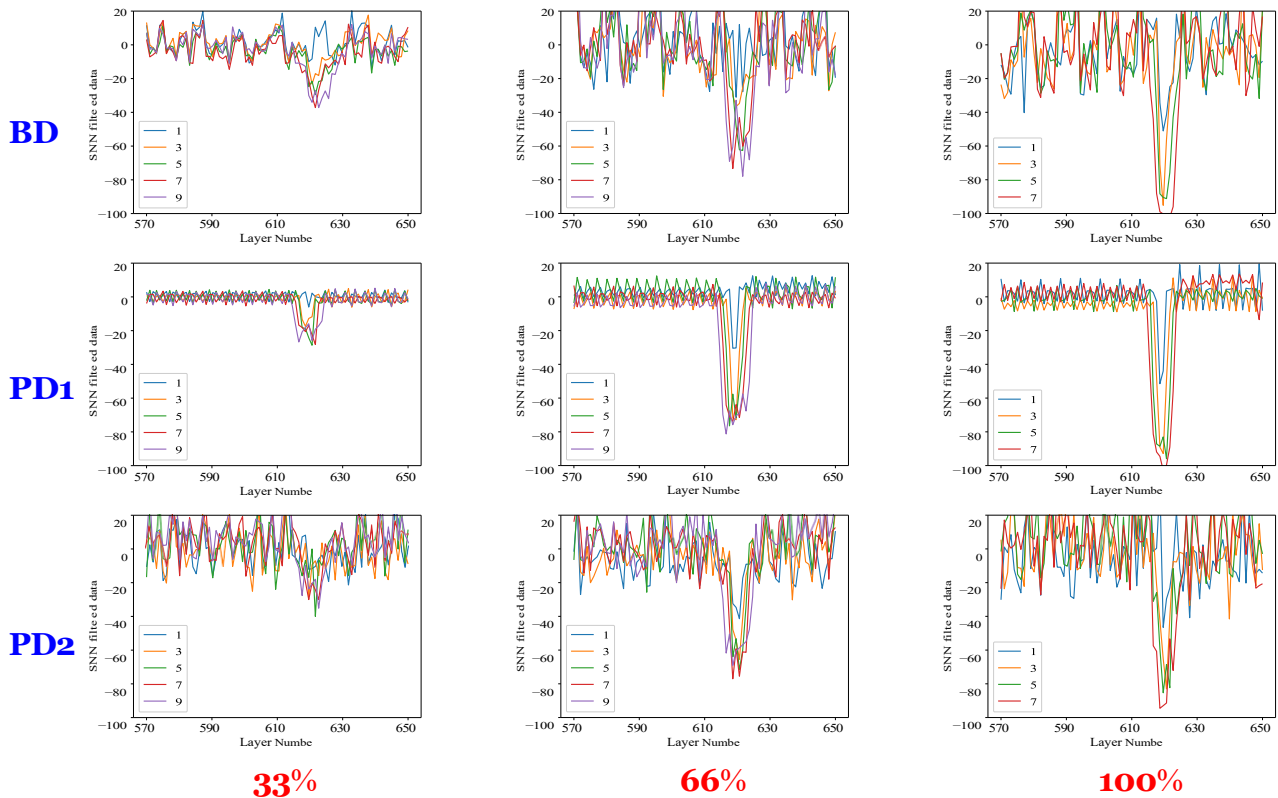

*3.1. Classification of Defect Samples using SNNs*

An SNN-based classifier is designed to classify the various defective samples. Since there is only one series of mean value per layer for a specific sample, the conventional training-test split is not possible in this case. However, attempt has been made to verify whether the SNN can distinguish each of the 14 samples, by causing a distinct output neuron to fire for every data sample. A cross-entropy loss function is used in this case, for the multi-class classification problem. The expression for this loss function is given in Equation 5.

$$L = -\frac{1}{N} \Sigma_{i=1}^{N} \Sigma_{j=1}^{C} y_{ij} \log(p_{ij}) \qquad (5)$$

,where $N$ is the total number of elements in the training batch, $C$ is the total number of classes, $y_{ij}$ is 1 if the $i^{th}$ element belongs to the $j^{th}$ class, otherwise 0, and $p_{ij}$ is the probability of the $j^{th}$ class obtained as the output of the SNN for the $i^{th}$ input. An increase in the value of the probability of the desired class, causes the log value to increase, and hence the loss $L$ to decrease. Thus for successful training, the cross entropy loss should decrease with training iterations.

Figure 10 shows the classification loss progression with training and the classification result. The classification accuracy obtained is 71.43%. This is because the classifier could not distinguish between 66% and 100% energy reduction scenarios for the respective layers in the sample.

Figure 10: (a)Change of cross-entropy loss with training, (b) Predicted sample versus target sample plot.

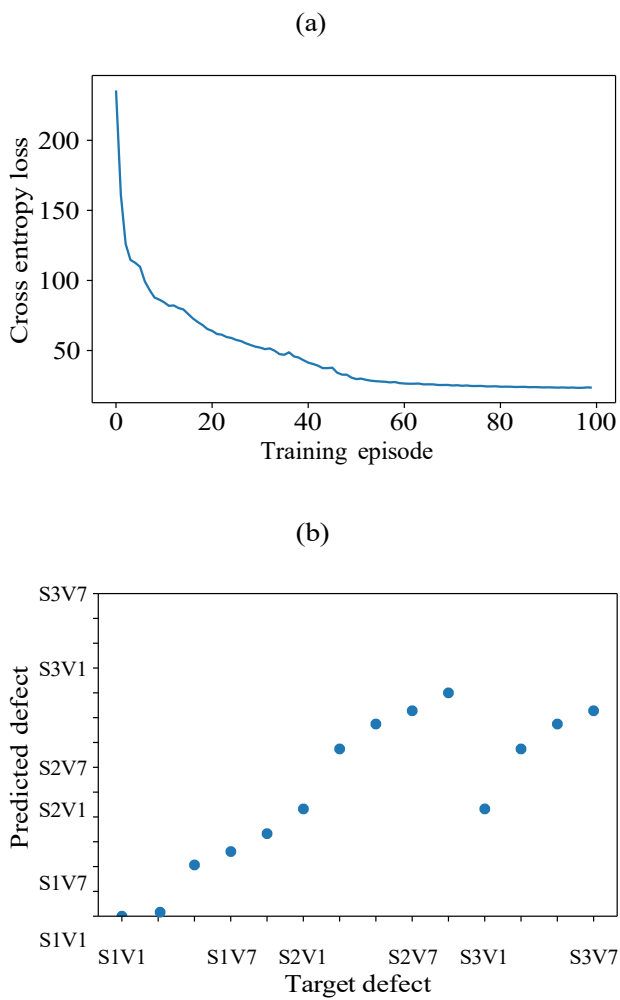

### 3.2. Energy comparison across hardware

Th reduction in computational energy consumption is a key consideration for neuromorphic hardware. Here, energy consumed is calculated from the energy per unit values as reported by works from Davies *et. al* [33, 32, 34]. The number of operations per inference are calculated from the network. Figure 11 shows the energy consumption estimates across neuromorphic and non-neuromorphic hardwares, and also compares the computational energy consumption in Loihi across various Ti-6Al-4V LPBF manufactured units. Quantitative estimates have been tabulated in Table 7. As expected, the energy consumption fluctuates with the samples as the firing rates change for every sample. Also, in this scenario, Loihi consumes fairly low amount of energy as compared to other hardware architectures, amounting to $0.821\mu J$.

Figure 11: (a)Energy consumption across CPU, GPU, FPGA, Loihi and SpiNNaker for 66% energy reduction across 7 layers, for the plasma radiations recorded by PD1 sensor(b) Shows the variation of Loihi's energy consumption with different samples, for the recordings of the PD1 sensor.

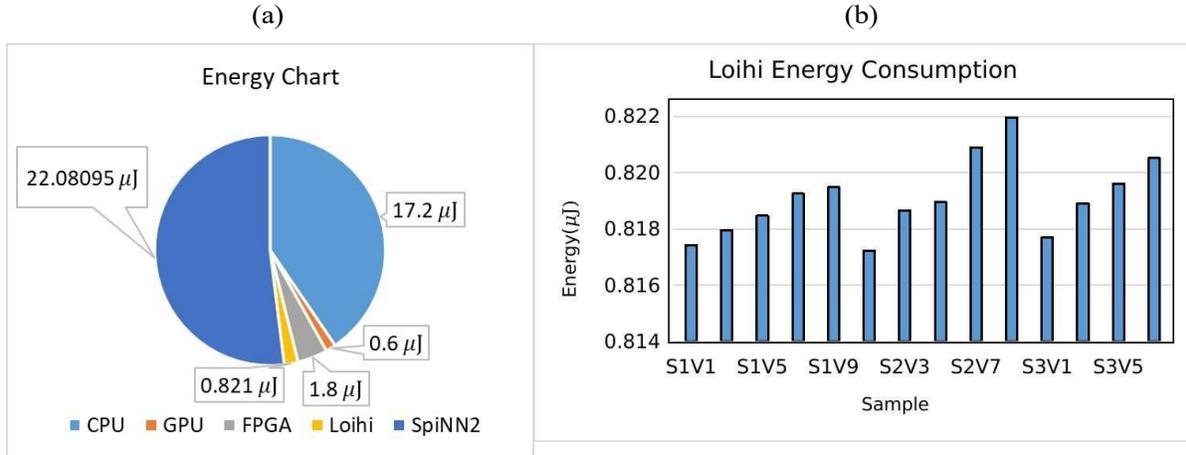

Table 7: Energy consumption across hardware in $\mu J/inference$

| Sample | CPU  | GPU | FPGA | Loihi | SpiNNaker2 |
|--------|------|-----|------|-------|------------|
| S1V3   | 17.2 | 0.6 | 1.8  | 0.818 | 22.0       |
| S1V7   | 17.2 | 0.6 | 1.8  | 0.819 | 22.1       |
| S2V3   | 17.2 | 0.6 | 1.8  | 0.819 | 22.0       |
| S2V7   | 17.2 | 0.6 | 1.8  | 0.821 | 22.1       |
| S3V3   | 17.2 | 0.6 | 1.8  | 0.819 | 22.0       |
| S3V7   | 17.2 | 0.6 | 1.8  | 0.821 | 22.1       |

## 4. Conclusion and Future Directions

This work demonstrates a first example of the applicability of the use of SNNs using the NEF framework to detect anomalies in LPBF printing process. To achieve this, sensor data from this additive manufacturing process is filtered and smoothened, and subsequently checked for anomalies. In addition, while allowing for low power, low latency computing capabilities suitable for edge applications. Sudden changes in the photodiode sensor signals are observed for various levels of print layer anomalies, corresponding to the reduction in laser power. These results are supported by implementations in FPGA and Loihi. There is statistically significant difference in measurements, even for lower levels of damage and the detection of presence, location and extent of damage. This implementation can be automatically carried out, even in the presence of noise which often mask such changes and may go visually undetected. It is seen that SNN provides a higher F1 score than Savitzky-Golay and Butterworth filtering techniques but provides a poorer F1 score than moving average or Gaussian techniques of filtering.

When compared across hardware, CPU shows the best performance as the noise levels are quite low for non-neuromorphic digital hardware. The same is the case for FPGA with a lower fluctuation than that of CPU. A hybrid CPU-FPGA approach filtering is applied at 3 stages (unlike 2 for CPU) and clipping through 2 ensembles (unlike 1 for CPU), resulting in a higher smoothing effect for the FPGA, making detection of anomalies easier. For Loihi, as is expected in the case of neuromorphic hardware, noise levels are higher even after filtering but the anomaly detection was still possible. Estimates of power consumption for neuromorphic and non-neuromorphic hardware shows that Loihi consumes significantly less power compared to CPU or FPGA.

The results also create a benchmark application for neuromorphic implementation for additive manufacturing and related anomalies. In future, such implementation and related investigations of performance around anomaly detection in real-time and for edge networks will be of interest and importance.

## Acknowledgements

This publication has emanated from research supported in part by a grant from Taighde Éireann – Research Ireland under Grant number 18/CRT/6049. For the purpose of Open Access, the author has applied a CC BY public copyright licence to any Author Accepted Manuscript version arising from this submission. Vikram Pakrashi and Vikram Pakrashi also acknowledges Research Ireland MaREI RC2302-2, NexSYs 21/SPP/3756 and Harmoni 22FFP-P11457.

## Disclosure Stateement

No potential conflict of interest was reported by the author(s).

## References

[1] Osama Abdulhameed, Abdulrahman Al-Ahmari, Wadea Ameen, and Syed Hammad Mian. Additive manufacturing: Challenges, trends, and applications. *Advances in Mechanical Engineering*, 11(2):1687814018822880, 2019.

[2] David L Bourell, Ming Leu, and David Rosen. *Roadmap for additive manufacturing: identifying the future of freeform processing*. University of Texas at Austin Labratory for Freeform Fabrication Advanced . . . , 2009.

[3] Darragh S. Egan and Denis P. Dowling. Influence of process parameters on the correlation between in-situ process monitoring data and the mechanical properties of ti-6al-4v non-stochastic cellular structures. *Additive Manufacturing*, 30:100890, 2019. ISSN 2214-8604. doi:https://doi.org/10.1016/j.addma.2019.100890. URL https://www.sciencedirect.com/science/article/pii/S2214860419307444.

[4] Mohammad Ghayoomi Mohammadi, Dalia Mahmoud, and Mohamed Elbestawi. On the application of machine learning for defect detection in l-pbf additive manufacturing. *Optics & Laser Technology*, 143:107338, 2021.

[5] Aoife C Doyle, Darragh S Egan, Caitríona M Ryan, Andrew C Parnell, and Denis P Dowling. Application of the stray statistical learning algorithm for the evaluation of in-situ process monitoring data during l-pbf additive manufacturing. *Procedia Manufacturing*, 54:250–256, 2021.

[6] Luke Scime and Jack Beuth. Anomaly detection and classification in a laser powder bed additive manufacturing process using a trained computer vision algorithm. *Additive Manufacturing*, 19:114–126, 2018.

[7] Luke Scime and Jack Beuth. A multi-scale convolutional neural network for autonomous anomaly detection and classification in a laser powder bed fusion additive manufacturing process. *Additive Manufacturing*, 24:273–286, 2018.

[8] Tayyaba Sahar, Muhammad Rauf, Ahmar Murtaza, Lehar Asip Khan, Hasan Ayub, Syed Muslim Jameel, and Inam Ul Ahad. Anomaly detection in laser powder bed fusion using machine learning: A review. *Results in Engineering*, 17:100803, 2023.

[9] Alvin Chen, Fotis Kopsaftopoulos, and Sandipan Mishra. An unsupervised online anomaly detection method for metal additive manufacturing processes via a statistical time-frequency domain algorithm. *Structural Health Monitoring*, 23(3):1926–1948, 2024.

[10] Mohamad Mahmoudi, Ahmed Aziz Ezzat, and Alaa Elwany. Layerwise anomaly detection in laser powder-bed fusion metal additive manufacturing. *Journal of Manufacturing Science and Engineering*, 141(3):031002, 2019.

[11] Catherine D Schuman, Shruti R Kulkarni, Maryam Parsa, J Parker Mitchell, Prasanna Date, and Bill Kay. Opportunities for neuromorphic computing algorithms and applications. *Nature Computational Science*, 2(1):10–19, 2022.

[12] Amar Shrestha, Haowen Fang, Zaidao Mei, Daniel Patrick Rider, Qing Wu, and Qinru Qiu. A survey on neuromorphic computing: Models and hardware. *IEEE Circuits and Systems Magazine*, 22(2):6–35, 2022.

[13] Felix Christian Bauer, Dylan Richard Muir, and Giacomo Indiveri. Real-time ultra-low power ecg anomaly detection using an event-driven neuromorphic processor. *IEEE Transactions on Biomedical Circuits and Systems*, 13(6):1575–1582, 2019. doi:10.1109/TBCAS.2019.2953001.

[14] Stefan Gerber, Marc Steiner, Giacomo Indiveri, Elisa Donati, et al. Neuromorphic implementation of ecg anomaly detection using delay chains. In *2022 IEEE Biomedical Circuits and Systems Conference (BioCAS)*, pages 369–373. IEEE, 2022.

[15] Guang Chen, Peigen Liu, Zhengfa Liu, Huajin Tang, Lin Hong, Jinhu Dong, Jörg Conradt, and Alois Knoll. Neuroaed: Towards efficient abnormal event detection in visual surveillance with neuromorphic vision sensor. *IEEE Transactions on Information Forensics and Security*, 16:923–936, 2020.

[16] Lakshmi Annamalai, Anirban Chakraborty, and Chetan Singh Thakur. Evan: Neuromorphic event-based sparse anomaly detection. *Frontiers in Neuroscience*, 15:699003, 2021.

[17] Yassine Jaoudi. Evaluating online learning anomaly detection on intel neuromorphic chip and memristor characterization tool. Master's thesis, University of Dayton, 2021.

[18] George Vathakkattil Joseph and Vikram Pakrashi. Spiking neural networks for structural health monitoring. *Sensors*, 22(23):9245, 2022.

[19] Ruimin Chen, Mohsen Imani, and Farhad Imani. Joint active search and neuromorphic computing for efficient data exploitation and monitoring in additive manufacturing. *Journal of manufacturing processes*, 71:743–752, 2021.

[20] Trevor Bekolay, James Bergstra, Eric Hunsberger, Travis DeWolf, Terrence C Stewart, Daniel Rasmussen, Xuan Choo, Aaron Russell Voelker, and Chris Eliasmith. Nengo: a python tool for building large-scale functional brain models. *Frontiers in neuroinformatics*, 7:48, 2014.

[21] Terrence C Stewart. A technical overview of the neural engineering framework. *University of Waterloo*, 110, 2012.

[22] Mike Davies, Andreas Wild, Garrick Orchard, Yulia Sandamirskaya, Gabriel A Fonseca Guerra, Prasad Joshi, Philipp Plank, and Sumedh R Risbud. Advancing neuromorphic computing with loihi: A survey of results and outlook. *Proceedings of the IEEE*, 109(5):911–934, 2021.

[23] Jesse Hagenaars, Federico Paredes-Vallés, and Guido De Croon. Self-supervised learning of event-based optical flow with spiking neural networks. *Advances in Neural Information Processing Systems*, 34:7167–7179, 2021.

[24] Vijay Shankaran Vivekanand, Shahin Hashemkhani, Shanmuga Venkatachalam, and Rajkumar Kubendran. Robot locomotion control using central pattern generator with non-linear bio-mimetic neurons. In *2023 9th International Conference on Automation, Robotics and Applications (ICARA)*, pages 102–106. IEEE, 2023.

[25] Chris Eliasmith and Charles H Anderson. *Neural engineering: Computation, representation, and dynamics in neurobiological systems*. MIT press, 2003.

[26] Aaron Russell Voelker. Dynamical systems in spiking neuromorphic hardware. 2019.

[27] Renishaw Plc. Renishaw: Infiniam spectral. URL https://www.renishaw.com/en/infiniam-spectral--42310?srsltid=AfmBOoryGpSfcq5RniZGVKZYzuiowqZBvDnn-kDt6A-_xyroYSaLrgZP.

[28] Thang Viet Huynh. Fpga-based acceleration for convolutional neural networks on pynq-z2. *International Journal Of Computing and Digital System*, 2021.

[29] Jonathan Sanderson and Syed Rafay Hasan. System integration of xilinx dpu and hdmi for real-time inference in pynq environment with image enhancement. In *2024 IEEE International Symposium on Circuits and Systems (ISCAS)*, pages 1–5. IEEE, 2024.

[30] S Madhava Prabhu and Seema Verma. A comprehensive survey on implementation of image processing algorithms using fpga. In *2020 5th IEEE International Conference on Recent Advances and Innovations in Engineering (ICRAIE)*, pages 1–6. IEEE, 2020.

[31] Chit-Kwan Lin, Andreas Wild, Gautham N Chinya, Yongqiang Cao, Mike Davies, Daniel M Lavery, and Hong Wang. Programming spiking neural networks on intel's loihi. *Computer*, 51(3):52–61, 2018.

[32] Mike Davies, Narayan Srinivasa, Tsung-Han Lin, Gautham Chinya, Yongqiang Cao, Sri Harsha Choday, Georgios Dimou, Prasad Joshi, Nabil Imam, Shweta Jain, et al. Loihi: A neuromorphic manycore processor with on-chip learning. *Ieee Micro*, 38(1):82–99, 2018.

[33] Brian Degnan, Bo Marr, and Jennifer Hasler. Assessing trends in performance per watt for signal processing applications. *IEEE Transactions on Very Large Scale Integration (VLSI) Systems*, 24(1):58–66, 2015.19

[34] Sebastian Höppner, Bernhard Vogginger, Yexin Yan, Andreas Dixius, Stefan Scholze, Johannes Partzsch, Felix Neumärker, Stephan Hartmann, Stefan Schiefer, Georg Ellguth, et al. Dynamic power management for neuromorphic many-core systems. *IEEE Transactions on Circuits and Systems I: Regular Papers*, 66(8):2973–2986, 2019.